\documentclass[aps,prd,reprint,nofootinbib, superscriptaddress]{revtex4-1}
\usepackage{amsmath,amssymb, amsthm,amstext}
\usepackage{natbib}
\usepackage{graphicx}
\usepackage{color}
\usepackage{array, enumerate}
\usepackage{bm}
\usepackage{multirow}
\usepackage[breaklinks,colorlinks,citecolor=blue]{hyperref}

\def\be{\begin{equation}}
\def\ee{\end{equation}}

\begin{document}
\title{Searching for Signatures of Dark matter-Dark Radiation Interaction \\
in Observations of Large-scale Structure}
\author{Zhen Pan}
\email{zhpan@ucdavis.edu}
\affiliation{Department of Physics, University of California, Davis, CA, 95616, USA}
\author{Manoj Kaplinghat}
\email{mkapling@uci.edu}
\affiliation{Department of Physics and Astronomy, University of California, Irvine, CA, 92697, USA}
\author{Lloyd Knox}
\email{lknox@ucdavis.edu}
\affiliation{Department of Physics, University of California, Davis, CA, 95616, USA}

\date{\today}

\begin{abstract}
    In this paper, we conduct a search in the latest large-scale structure
    measurements for signatures of the dark matter-dark radiation interaction
    proposed by Buen-Abad et al. (2015). We show that prior claims of a
    detection of this interaction rely on a use of the SZ cluster mass function
    that ignores uncertainty in the mass-observable relationship. Including this
    uncertainty we find that the inferred level of interaction remains consistent
    with the data, but so does zero interaction; i.e., there is no longer a
    detection. We also point out that inference of the shape and amplitude of
    the matter power spectrum from Ly$\alpha$ forest measurements is highly
    inconsistent with the predictions of the $\Lambda$CDM model conditioned on
    Planck CMB temperature, polarization, and lensing power spectra, and that
    the dark matter-dark radiation model can restore that consistency. We also
    phenomenologically generalize the model of Buen-Abad et al. (2015) to allow for interaction
    rates with different scalings with temperature, and find that the original
    scaling is preferred by the data.
\end{abstract}

\maketitle

\section{Introduction}
Dark matter is an essential component of the standard $\Lambda$CDM cosmology,
whose existence has been established from many cosmological and astrophysical
lines of evidence \cite[see e.g.][for a brief summary]{Sanders2010,Bertone2016,DeSwart2017}.
On the other hand, increasingly sensitive efforts at direct
detection of canonical candidates such as WIMPs and axions
have only resulted in upper limits \cite{Aprile:2017iyp,Sloan:2016aub}. The lack of direct detection signatures implies that the dark matter is weakly coupled to the standard model, but it does not preclude a large coupling to a hidden sector. The idea of hidden sector dark matter has broadened the experimental search possibilities, while retaining some of the virtues of WIMP models such as concrete thermal and non-thermal production mechanisms \cite[e.g.,][]{MarchRussell:2008yu,Feng:2008ya,Feng:2008mu,Cheung:2010gj} and opening up new cosmological signatures \cite[e.g.,][]{Feng:2009mn,Cyr-Racine2013}.

The richer phenomenology expands the possible ways in which dark matter properties
may be revealed through observations of the large-scale structure (LSS) of the universe.
Precision measurement of the Cosmic Microwave Background (CMB) temperature and
polarization, as well as large-scale photometric and spectroscopic galaxy surveys
can be used to detect the influence of non-trivial dark matter properties or to limit
them \cite{s42016}. Indeed, the $\sigma_8$ tension in $\Lambda$CDM
cosmology, that LSS surveys yield lower $\sigma_8$ values than that derived from
CMB observations, is potentially due to non-trivial dark matter interactions and
has also served to renew interest in exploration of broader classes of dark matter
models \cite[e.g.][]{Kaplan2009,Kaplan2011,Cyr-Racine2013,
Diamanti2013,Archidiacono2015,Chacko2016,Cyr-Racine2016,
Prilepina2016,Buen-Abad2015,Lesgourgues2015,Ko2017, Ko2017b,Tang2016,Tang2017,
Krall2017,Raveri2017,Buen-Abad2017,DiValentino2017c}.

 In this paper, we focus on the non-Abelian dark sector scenario
 proposed by \cite{Buen-Abad2015},
 where the dark matter is a Dirac fermion that transforms under a non-Abelian gauge group, and the dark
 radiation is the associated gauge field, a massless ``dark gluon''.
 The strong self-interaction of the dark radiation makes it behave like a fluid,
 instead of a free-streaming species.
 The interaction between dark matter and the dark
 radiation fluid (dm-drf) acts to suppress the matter power spectrum,
 improving agreement with lower $\sigma_8$ values
 derived from LSS measurements \cite{PlanckCollaborationXV2015a,Heymans2013,%
  Hildebrandt2016, Kohlinger2017, DESCollaboration2017a, DESWL2017,Benson2013,%
  PlanckCollaborationXX2014, PlanckCollaborationXXIV2015a}.
 Some previous works \cite{Lesgourgues2015,Krall2017,Buen-Abad2017} show that
 the dm-drf interaction is detected at about $3\sigma$ confidence level jointly using
 \textsl{Planck} CMB and  LSS measurements,
 including \textsl{Planck} CMB lensing \cite{PlanckCollaborationXV2015a},
 CFHTLens weak lensing (WL) \cite{Heymans2013} and
 \textsl{Planck} Sunyaev-Zeldovich (SZ) cluster counters \cite{PlanckCollaborationXX2014, PlanckCollaborationXXIV2015a}.
The authors of \cite{Lesgourgues2015} emphasize the possibility for these models
to alleviate the $H_0$ tension \cite{Riess2016, Follin2017} as well.

Throughout this paper, we examine this non-Abelian dark sector model
with these LSS measurements one by one.
We critically review the analyses done previously
and identify which data set is driving the previously claimed dm-drf interaction;
we also show that the latest inferences of the matter power spectrum
from Lyman-$\alpha$ (Ly$\alpha$) forest data are highly inconsistent
with the \textsl{Planck} CMB data, assuming $\Lambda$CDM;
and finally we show that these data sets can be made
consistent by allowing for this dark sector model.

We find that \textsl{Planck} SZ is essential for the
claimed detection of the dm-drf interaction in previous analyses.
But the SZ cluster constraint is limited by a large uncertainty in the cluster mass
scale determination, which is usually parametrized by a mass bias parameter $b$.
The bias parameter $b$ itself is constrained by several different analyses of the gravitational lensing induced by SZ galaxy clusters including two using the distorted shapes of background galaxies (``Weighing the Giants'' \cite{VonderLinden2012} and the Canadian Cluster Comparison Project  \cite{Hoekstra2015})
and one using distortions of the CMB \cite{Melin2015}. The inferred value of $\sigma_8$ from the observations of the SZ clusters depends sensitively on the mass estimates of the clusters and therefore on the mass bias parameter $b$. While previous dm-drf analyses effectively assumed zero uncertainty in $b$, we find that including an uncertainty based on any of the above inferences of $b$, the claimed detection turns into an upper limit.

Compared with the other LSS measurements mentioned above,
the Ly$\alpha$ forest power spectrum is sensitive to the matter power spectrum
at smaller scale $k\sim {\rm Mpc}^{-1}$, a scale that is more sensitive to the strength of the interaction in the dm-drf model. We find that the matter power spectrum derived from the
latest Ly$\alpha$ forest data \cite{Palanque-Delabrouille2013,Palanque-Delabrouille2015c,Palanque-Delabrouille2015b} is much steeper than that derived from \textsl{Planck} CMB data,
assuming $\Lambda$CDM. Finally, we examine whether this Ly$\alpha$-CMB tension
can be resolved by the dm-drf interaction.

The paper is organized as follows.
In Section \ref{sec:impacts},
we briefly introduce the non-Abelian dark sector model
and its impacts on the CMB power spectra and the matter power spectrum.
In Section \ref{sec:sz},
we show that using SZ data with the mass bias parameter fixed or varying makes a
huge difference in the constraints of cosmological parameters.
In Section \ref{sec:lyman}, we point out the Ly$\alpha$-CMB tension
in the $\Lambda$CDM cosmology,
and that the joint dataset favors a non-zero dm-drf interaction.
In Section \ref{sec:general}, we extend the exploration to
more general dm-drf interaction models characterized by interaction rates
scaling with temperature in different ways, and we also examine these models against
CMB data and LSS measurements.
We provide a summary in Section \ref{sec:summary}.

\section{Canonical DM-DRF Interaction Model}
\label{sec:impacts}

Following Ref. \cite{Buen-Abad2015},
we use $\Gamma$ for the momentum transfer rate
(i.e., time scale for momentum of dark matter particles
to change by ${\cal O}(1)$) due to the dm-drf scattering.
This momentum transfer leads to a drag force on the non-relativistic
dark matter particles such that
$\dot {\vec {v}}_{\rm dm} = a \Gamma (\vec v_{\rm drf} - \vec v_{\rm dm})$,
where $a$ is the scale factor and throughout this paper
dots denote conformal time derivatives.
In terms of physical quanties, $\Gamma$ is approximately
\be
\label{eq:gamma}
\Gamma\simeq (T_{\rm drf}/m_{\rm dm}) n_{\rm drf} \sigma_{\rm dm-drf},
\ee
where $m_{\rm dm}$ is the mass of dark matter particles,
$\sigma_{\rm dm-drf}$ is the cross section of dm-drf scattering, and
$T_{\rm drf}$ and $n_{\rm drf}$ are the temperature and the number density
of dark radiation, respectively.
For the non-Abelian dark sector model proposed by \cite{Buen-Abad2015},
$\sigma_{\rm dm-drf}\propto T_{\rm drf}^{-2}$,
thus we can write
$\Gamma = \Gamma_0(T/T_0)^2$, where $\Gamma_0$ denotes the velocity change rate today.
With this parametrization, the evolution equations of dark
matter density and velocity perturbations, $\delta_{\rm dm}$ and $\theta_{\rm dm}$,
and dark radiation density and velocity perturbations,
$\delta_{\rm drf}$ and $\theta_{\rm drf}$, are written as \cite{Lesgourgues2015}
\be
\begin{aligned}
    \dot \delta_{\rm dm} &= -\theta_{\rm dm} + 3 \dot \phi, \\
    \dot \theta_{\rm dm} &= \frac{\dot a}{a} \theta_{\rm dm} + k^2\psi + a\Gamma (\theta_{\rm drf} - \theta_{\rm dm}),\\
    \dot \delta_{\rm drf} &= -\frac{4}{3}\theta_{\rm drf} + 4 \dot \phi, \\
    \dot \theta_{\rm drf} &= k^2\frac{\delta_{\rm dr}}{4} + k^2\psi + \frac{3\rho_{\rm dm}}{4\rho_{\rm drf}}a\Gamma (\theta_{\rm dm} - \theta_{\rm drf}).
\end{aligned}
\ee
in the Conformal Newtonian gauge,
where
$\rho_{\rm dm}$ and $\rho_{\rm drf}$ are the average densities
of dark matter and dark radiation, respectively;
$\psi$ and $\phi$ are the Newtonian potential and the perturbation to
the spatial curvature, respectively.\footnote{The perturbation evolution equations in
the Boltzmann code {\tt CAMB}
are written in the Synchronous gauge. In our modfied version {\tt CAMB},
we keep a tiny amount of non-interacting dark matter to carry the Synchronous gauge.}

In the remainder of this section, we qualitatively
explain the impacts of dm-drf interaction
on cosmological observables,
by comparing the matter power spectra (Figure \ref{fig:mpk})
and CMB power spectra (Figure \ref{fig:TTEE})
of three cosmologies with  different parameters
$\{N_\nu, N_{\rm drf}, \Gamma_0\}$ and same other
parameters (see Table \ref{table:3models}).
Similar numerical comparisons were also given in previous
works \cite{Lesgourgues2015,Buen-Abad2017}, and here we focus on
connecting the impacts on observables with underlying physics.

\begin{table}
    \centering
\begin{tabular}{ c|c c  c}
 & $N_\nu$  & $N_{\rm drf}$ & $10^7\Gamma_0(\rm Mpc^{-1})$ \\
  \hline
 Model 1 & 3.546 & 0.0 & 0 \\
 Model 2 & 3.046 & 0.5 & 0 \\
 Model 3 & 3.046 & 0.5 & 2 \\
 \hline
\end{tabular}
\caption{\label{table:3models}
Three models used for clarifying the impact of dark radiation fluid on cosmological
observables, with different parameters $\{N_\nu, N_{\rm drf}, \Gamma_0\}$
and same $\Lambda$CDM parameters $\omega_{\rm b} = 0.02253, \omega_{\rm dm}= 0.1122,
A_{\rm s}=2.42\times10^{-9} , n_{\rm s}=0.967, \tau=0.0845, H_0= 70.4$ km/s/Mpc. }
\end{table}

\subsection{LSS}

\begin{figure}
\includegraphics[scale=0.55]{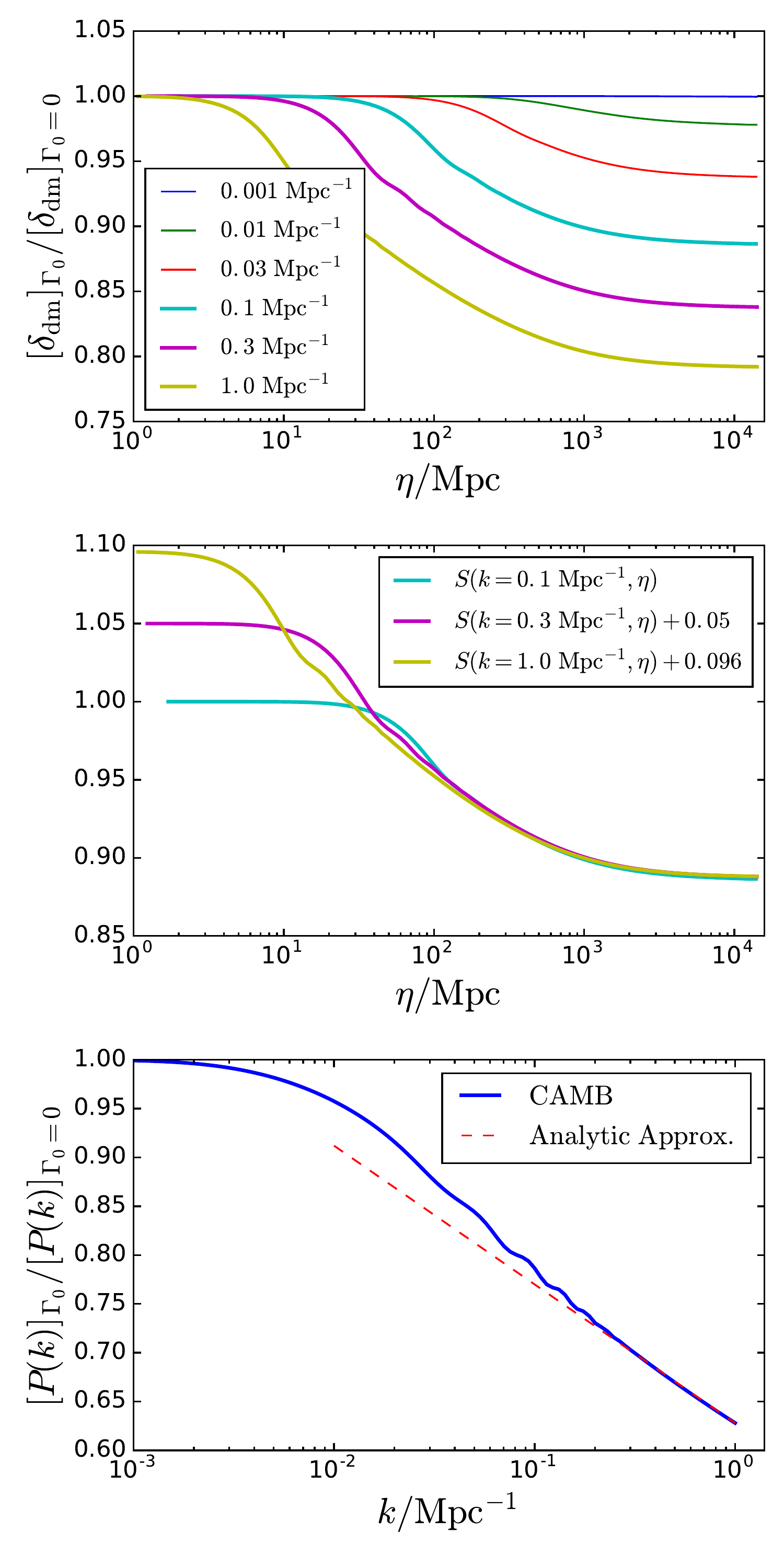}
\caption{\label{fig:mpk}
Upper Panel: the evolution of dark matter over-density $\delta_{\rm dm}$ suppression
for different $k$ modes.
Middle Panel: the ``self-similar" behavior of the over-density suppression,
where we displace the suppression of modes $k=0.3 \ {\rm Mpc}^{-1}$
and $k=1.0 \ {\rm Mpc}^{-1}$ by $0.05$ and $0.096$ respectively.
Lower Panel: the dm-drf interaction induced matter power spectrum suppression today,
where the dashed line is an analytic fit in the form of Equation (\ref{eq:fit}).}
\end{figure}

In the standard $\Lambda$CDM cosmology, the
dark matter over-density $\delta_{\rm dm}$ grows
logarithmically in the radiation-dominated era,
and grows linearly in the matter-dominated era \cite[e.g.,][]{Hu1995}.
A small dm-drf interaction does not remove these
growth modes, instead it decreases the corresponding growth rates.
We define the over-density suppression function
$\mathcal S(k,\eta) \equiv [\delta_{\rm dm}]_{\Gamma_0>0}/[\delta_{\rm dm}]_{\Gamma_0=0}$,
and plot $\mathcal S(k,\eta) = [\delta_{\rm dm}]_{\rm Model \ 3}/[\delta_{\rm dm}]_{\rm Model\ 2}$
(Table \ref{table:3models})
for different $k$ modes in Figure \ref{fig:mpk}.
We see that $\mathcal S(k,\eta)$ shows a ``self-similar" behavior
for small-scale modes $k\gg k_{\rm eq}$:
approximately,
\be
\begin{aligned}
    &\mathcal S(k,\eta) = 1 \qquad \qquad \qquad\,   (k\eta \lesssim 1),\\
    &\mathcal S(k,\eta) \simeq 1 - A  \log(k\eta) \quad ( 1 \lesssim k\eta \lesssim k \eta_{\rm eq}), \\
\end{aligned}
\ee
and after  radiation-matter equality, the  evolution of $\mathcal S(k,\eta)$
is similar for all different modes, where the suppression at
radiation-matter transition and that  today differ by a constant number,
\be
    \mathcal S(k, \eta_{\rm eq}) -\mathcal S(k,\eta_0)  \simeq B,
\ee
where $1/k_{\rm eq}\approx\eta_{\rm eq} \approx 100$ Mpc,
$\eta_0$ is the conformal time today, $A$ and $B$ are numbers
independent of mode $k$ and time $\eta$
(for the example shown in Figure \ref{fig:mpk}, $A\approx 0.04$ and $B\approx 0.05$).
The self-similar behavior for modes $k\gg k_{\rm eq}$
originates from the fact $\Gamma/H$ is a constant in the radiation-dominated era,
and therefore introduces no new timescale or length scale.

With the approximations above, we can estimate the power spectrum suppression today as
\be
\label{eq:fit}
\begin{aligned}
\frac{\left[P(k)\right]_{\Gamma_0>0}}{\left[P(k)\right]_{\Gamma_0=0}}
&= (1 - A  \log(k/k_{\rm eq}) -B)^2  \\
&\approx 1-2B - 2A  \log(k/k_{\rm eq}),
\end{aligned}
\ee
where we have ignored quadratic terms in the second line.
This estimate explains the dm-drf interaction induced
logarithmic suppression in the matter power spectrum $P(k)$
for modes $k\gg k_{\rm eq}$
(see Figure \ref{fig:mpk} for the matter power suppression computed from {\tt CAMB} and
the logarithmic  fit).

\subsection{CMB}
\label{subsec:impacts_cmb}

\begin{figure*}
\includegraphics[scale=0.3]{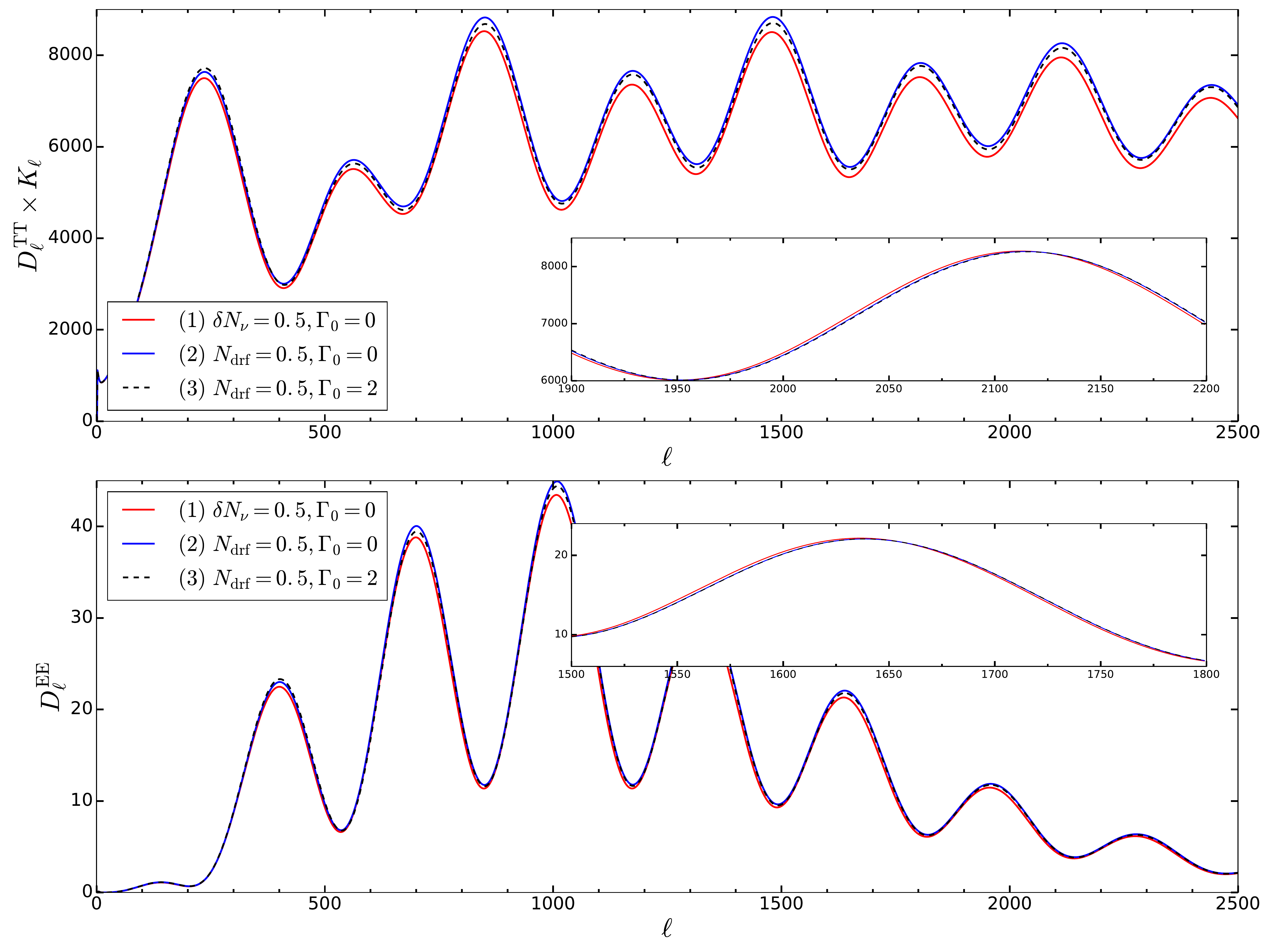}
\caption{Comparison of TT and EE spectra of the three models listed in Table \ref{table:3models},
where in the upper panel we plot the TT spectra with damping effect largely removed by
multiplying a factor $K_\ell = \exp\left\{2\times(\ell/1267)^{1.18}\right\}$,
in the lower panel we plot the EE spectra, and in the two inset plots,
we normalize the spectra amplitudes to allow one to see the impact of the very small shift in peak locations
induced by free-streaming species as done in \cite{Follin2015a}. In these insets the red curve
(the model with additional freestreaming neutrinos) is slightly
shifted to the left relative to the dashed line and blue line which overlap each other.}
\label{fig:TTEE}
\end{figure*}

The imprint of the dm-drf interaction on the CMB power spectra is much more subtle
as shown in Figure \ref{fig:TTEE}.
Comparison of Models 1 and 2 confirms the signatures of free-streaming neutrinos in the CMB spectra: namely
power suppression and a (very small) shift in acoustic peak locations \cite{Hu1995,Hu1996d,Hu1997d,Bashinsky2004a,Bashinsky2007,
Follin2015a, Baumann2016, Pan2016}.
Comparison of Models 2 and 3 shows that the dm-drf interaction very slightly
increases the amplitude of modes $\ell \lesssim 500$
and decreases that of modes $\ell \gtrsim 500$.

For modes $\ell \lesssim 500$, the increased amplitude
can be explained by the near-resonant driving of the baryon-photon
fluid perturbation amplitude by gravitational potential decay as
modes enter the horizon \cite{Hu1995,Hu1996d,Hu1997d}. The resistance
to dark matter free fall from the dm-drf interaction contributes to
gravitational potential decay, at least on scales large enough that,
at the time of horizon crossing, the dark matter contributes a
significant fraction of the total energy density.

For modes $\ell \gtrsim 500$, instead of an enhancement, we see
instead a very small suppression of power.
 and a small suppression of even-odd
peak height difference due to the dm-drf interaction.
The extra potential
decay arising from the interaction changes the photon overdensity in two ways:
amplitude suppression and baryon loading alleviation.
At these small scales, we numerically find that the extra
potential decay leads to a nearly uniform suppression of the
photon perturbation amplitude in a low-baryon cosmology.
We also find that the baryon loading effect is weaker in Model 3 than in Model 2.
The two changes (amplitude suppression and baryon loading alleviaion) add up constructively for the odd extrema ($kr_{\rm s, *}
= 3\pi, 5\pi, 7\pi$) and destructively for the even extrema ($kr_{\rm s, *}
= 2\pi, 4\pi, 6\pi$).\footnote{In fact,
\textsl{Planck} CMB data is sensitive to the small suppression of the
odd-even peak height difference. Our MCMC results show that the dm-drf model prefers
a higher $\omega_{\rm b}$ than in the $\Lambda$CDM model.}

To summarize, the dm-drf interaction has a much smaller impact on the CMB
power spectra than on the matter power spectrum. The impact on the matter
power spectrum arises from interactions in the radiation-dominated era when
the dark radiation has more inertia than the dark matter. The impact on the
CMB power spectrum is through the impact on the dark matter evolution.
On small scales,
where the impact on dark matter is sizeable, the dark matter contribution to
the gravitational potential at the time of horizon crossing is very small and
thus the net impact on the photon
distribution is small.

\section{Parameter constraints from LSS data}
\label{sec:sz}
In this section, we first briefly review previous analyses of the implications of cosmological
data for the extension of $\Lambda$CDM to include the dm-drf interaction model, identify \textsl{Planck} SZ as the major driver for the previously claimed detection of the dm-drf interaction,
and redo the analysis with a treatment of uncertainties in the SZ-mass observable relationship.

\subsection{Previous Analyses}

In previous analyses \cite[e.g.][]{Lesgourgues2015, Krall2017,Buen-Abad2017}, \textsl{Planck} CMB and LSS measurements, including
\textsl{Planck} CMB Lensing \cite{PlanckCollaborationXV2015a},
CFHTLens \cite{Heymans2013} and
\textsl{Planck} SZ \cite{PlanckCollaborationXX2014, PlanckCollaborationXXIV2015a},
\begin{subequations}
\begin{align}
    \sigma_8(\Omega_{\rm m}/ 0.27)^{0.25} &= 0.820\pm 0.029 \ [{\rm CMB \ Lensing}], \label{eq:lss1}\\
    \sigma_8(\Omega_{\rm m}/ 0.27)^{0.46} &= 0.774\pm 0.040 \ [{\rm CFHTLens}], \label{eq:lss2}\\
    \sigma_8(\Omega_{\rm m}/ 0.27)^{0.30} &= 0.782\pm 0.010 \ [{\rm Planck \ SZ}]. \label{eq:lss3}
\end{align}
\end{subequations}
were used to constrain the dm-drf model, and the dm-drf interaction was
detected at $3\sigma$ confidence level.

\begin{figure}
\includegraphics[scale=0.55]{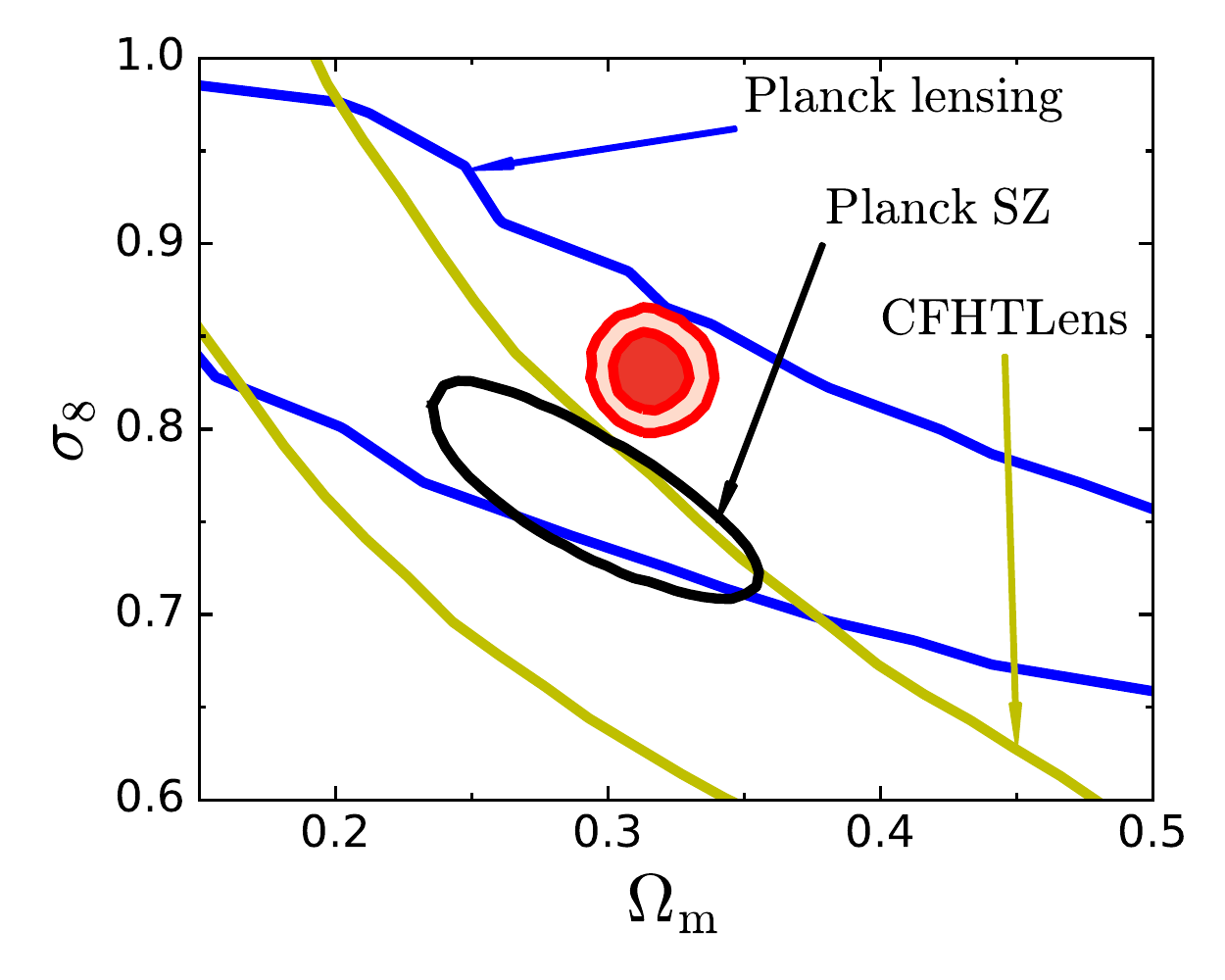}
\caption{\label{fig:tension}
The $\sigma_8$ tension in the $\Lambda$CDM cosmology,
where the red filled contours ($1\sigma$ and $2\sigma$) are derived from
\textsl{Planck} 2015 temperature and polarization,
and the unfilled contours corresponding to the
three LSS measurements are given at $2\sigma$ level,
where the SZ contour is the constraint fixing
the mass bias parameter as  the baseline value $1-b=0.8$.}
\end{figure}

To figure out which dataset is essential to the claimed  dm-drf
detection, we show the $\sigma_8$ tension of the $\Lambda$CDM cosmology in Figure \ref{fig:tension},
which clearly shows that the SZ-CMB tension is the strongest.
This finding also suggests that \textsl{Planck} SZ is
driving the detection.

The cosmological implications of the Planck SZ cluster counts depends on assumptions about the relationship between SZ flux and cluster mass \cite[e.g.,][]{PlanckCollaborationXXIV2015a}. This is usually expressed as uncertainty in the hydrostatic mass bias parameter $b$ where
$M_{\rm X} = (1-b) M_{500}$, $M_{\rm X}$ is a mass proxy derived
from observed SZ flux with an assumption of hydrostatic equilibrium, and
$M_{500}$ is the true cluster halo mass (see
\cite{PlanckCollaborationXX2014, PlanckCollaborationXXIV2015a} for more details).
The bias parameter $b$ itself is not well known today, e.g.,
constraints derived from gravitational shear mass measurements
Weighing the Giants (WtG) \cite{VonderLinden2012},
from Canadian Cluster Comparison Project (CCCP) \cite{Hoekstra2015},
and  from CMB Lensing (Lens) \cite{Melin2015,Hurier2017} listed as follows
show significant uncertainties:
\begin{subequations}
\begin{align}
   1-b &= 0.688\pm 0.072 \ [{\rm WtG}] \label{eq:wtg},\\
   1-b &= 0.780\pm 0.092 \ [{\rm CCCP}] \label{eq:cccp},\\
   1-b &= 0.74\pm 0.07\phantom{xx\, } [{\rm Lens}]. \label{eq:lens}
\end{align}
\end{subequations}
As shown in \cite{PlanckCollaborationXXIV2015a}, the $\sigma_8$ constraint derived from
SZ cluster counts is sensitive to the prior used:
the WtG prior almost eliminates the $\sigma_8$ tension between SZ and \textsl{Planck} CMB,
while the CCCP prior remains in noticeable tension.
In addition, a reference model fixing the bias parameter as the baseline value, $1-b=0.8$,
was also investigated in the \textsl{Planck} SZ analysis \cite{PlanckCollaborationXX2014},
which yields the $\sigma_8$ constraint of Equation (\ref{eq:lss3}),
in tension with that derived from \textsl{Planck} CMB at
$\gtrsim 3\sigma$ confidence level (see also Figure \ref{fig:tension}).

In previous analyses, the $\sigma_8$ constraint of Equation (\ref{eq:lss3})
was usually used as an approximation to the full \textsl{Planck} SZ data.
It is natural to ask whether it is valid to fix the bias parameter as the baseline value
in constraining the dm-drf interaction model, considering the large uncertainty
in the bias parameter, the sensitive dependence of the $\sigma_8$ constraint
on the bias parameter and the mild tension between the WtG constraint and the baseline value (see \cite{Salvati2017a} for a summary of recent bias parameter inferences).
We discuss this next.

\subsection{Anaysis with SZ data: the impact of the mass bias parameter}
To highlight the impact of the uncertainty in the SZ cluster counts
on the model parameter constraints,
we use both CMB and SZ data with the mass bias parameter fixed or varying,
and compare the resulting constraints. For CMB data, we use
\textsl{Planck} 2015 CMB temperature and polarization
data TTTEEE + lowTEB \cite{PlanckCollaborationXIII2015} ({\tt PlanckTP}).
For SZ data with varying mass bias parameter, we use
\textsl{Planck} 2015 SZ cluster counts data ({\tt PlanckSZ}) with the CCCP prior,
while for SZ data with fixed mass bias parameter, we
use the single data point of Equation (\ref{eq:lss3}),
as done in previous analyses.

\begin{figure}
\includegraphics[scale=0.42]{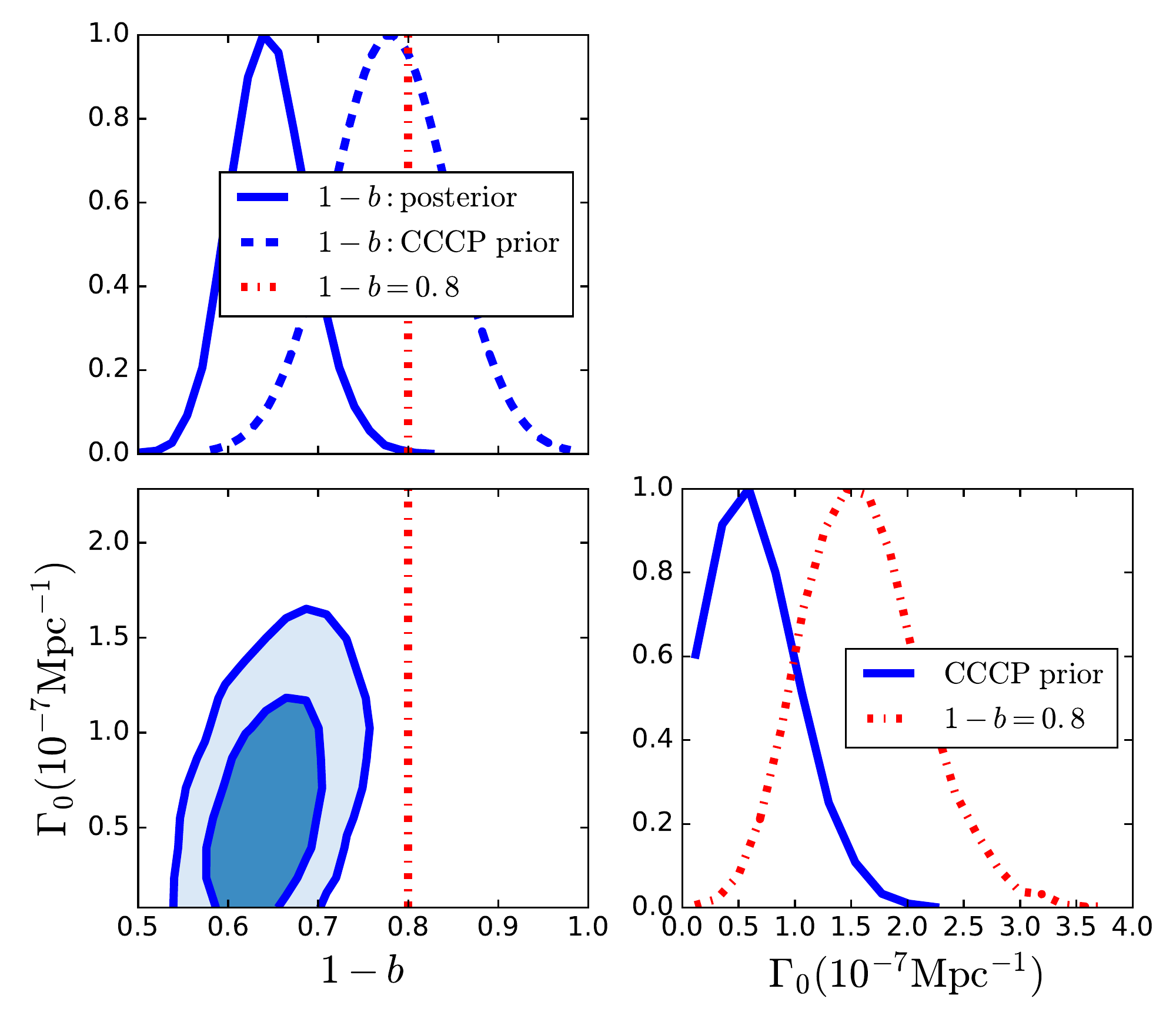}
\caption{\label{fig:Gamma0}
The MCMC results for the canonical dm-drf model using joint dataset {\tt PlanckTP+PlanckSZ}.
Upper Left Panel: the comparison of the CCCP prior and the resulting posterior with
the baseline value $1-b=0.8$. Lower Left Panel: the posterior contour of
$1-b$ vs. $\Gamma_0$. Lower Right Panel:
the marginalized posteriors of $\Gamma_0 $ with the bias parameter fixed or varying.}
\end{figure}

We use  {\tt CosmoMC} to run MCMC chains,
with flat priors on $N_{\rm drf}\geq 0.07$ and $\Gamma_0 \geq 0$,
and {\tt CosmoMC} default priors for $\Lambda$CDM parameters and other nuisance parameters.
We use the  Raferty and Lewis statistic $R-1\leq0.02$ as the convergence criterion,
and we summarize the MCMC results in Figure \ref{fig:Gamma0} and Table \ref{table:mcmc}.

\begin{table*}
    \centering
\begin{tabular}{ c| c | c | c | c }
 \multirow{2}{*}{Dataset} &
  \multirow{2}{*}{\tt PlanckTP} &
   \multicolumn{2}{c|}{\tt PlanckTP+PlanckSZ} &
  \multirow{2}{*}{\tt PlanckTP+Lensing+DES} \\
   &   & $1-b = 0.78\pm0.092$ & $\phantom{xxx} 1-b=0.8 \phantom{xxx}$ & \\
   \hline
 $1-b$ & & $0.647\pm 0.044$   &     & \\
 $\Gamma_0(10^{-7}{\rm Mpc}^{-1})$  & $<1.28$  & $<1.36$  &  $1.61\pm 0.54$ & $<1.43$\\
 $N_{\rm drf}$ & $< 0.57$  & $< 0.62$  & $< 0.64$ & $<0.67$\\
 $\sigma_8$ & $0.817\pm 0.022$ & $0.807\pm0.019$ & $0.758\pm0.015$ &$0.800\pm 0.016$ \\
 \hline
\end{tabular}
\caption{\label{table:mcmc}Constraints on the  dm-drf model parameters
using datasets {\tt PlanckTP}, {\tt PlanckSZ}, {\tt Lensing} and {\tt DES},
where the uncertainties are  $1\sigma$ values,
and the upper limits are given at $2\sigma$ confidence level.}
\end{table*}

Similar to previous works, we obtain a  $3\sigma$ detection of $\Gamma_0$
jointly using {\tt PlanckTP} and {\tt PlanckSZ} fixing the bias parameter as $1-b=0.8$,
but the detection disappears if we let the bias parameter vary and impose
the CCCP prior. In the latter case,
the posterior of the mass bias parameter turns out to converge at $1-b= 0.647\pm 0.044$,
which is about $3 \sigma$ lower than the baseline value $0.8$ and $2$
times tighter than the CCCP prior imposed,
due to the overriding \textsl{Planck} CMB preference for lower $1-b$ (higher $\sigma_8$);
and the detection of $\Gamma_0$ is gone due to the tight positive correlation
between $1-b$ and $\Gamma_0$. But the degeneracy of $1-b$ and $\Gamma_0$ breaks down
at $10^7\Gamma_0/{\rm Mpc^{-1}}\sim 1.5$ since \textsl{Planck} CMB power spectra
disfavor large $\Gamma_0$ (see Table \ref{table:mcmc}).

We also checked the approximation of using the single data point of Equation (\ref{eq:lss3}) rather than the full SZ likelihood. Equation (\ref{eq:lss3}) follows from the full SZ likelihood given the $\Lambda$CDM model and that $1-b=0.8$ with no uncertainty. We find the approximation works well. We find very similar constraints on the dm-drf interaction model parameters whether we use the full SZ likelihood (and $1-b = 0.8$) or approixmate it with Equation (\ref{eq:lss3}).
Both of them result in $\sim 3\sigma$ detection of the dm-drf interaction,
with tiny differences in the mean values and the uncertainties,
which do not affect our conclusion. We therefore do not distinguish the two cases in this paper.

From Table \ref{table:mcmc}, we also see that {\tt PlanckSZ} with the CCCP prior
is not highly constraining; adding it to {\tt PlanckTP}
only slightly increases the upper limits of $\Gamma_0$ and $N_{\rm drf}$.
It is clear that the other two priors would lead to even less of a $\Gamma_0$ detection,
which can be verified by the fact that the tension of the $1-b$ posterior with the
CCCP prior is greater than its tension with the WtG/Lens prior.

\subsection{Analysis with only CMB Lensing and DES data}

Since the SZ data (with the bias parameter allowed to float) is not highly constraining, we drop it from further consideration as we
we examine the dm-drf model with  {\tt PlanckTP}  and  the following two LSS datasets:

(1) {\tt Lensing}:  \textsl{Planck} 2015 lensing data \cite{PlanckCollaborationXV2015a}.

(2) {\tt DES}: $\sigma_8(\Omega_{\rm m}/0.3)^{0.5} = 0.789 \pm 0.026$,
    which is derived from the Dark Energy Survey (DES) first-year cosmic shear data \cite{DESWL2017}
    and is a slightly tighter constraint than that derived from CFHTLens (Eq.\ref{eq:lss2})
    or from KiDS-450 \cite{Hildebrandt2016a}.
    Strictly
    speaking, we should use the DES likelihood with all the relevant nuisance parameters
    (e.g. the intrinsic alignment of galaxies) varying, instead of using this single data point.
    But the likelihood code is not publicly available, and as we will see later, we find no
    detection of the dm-drf interaction. Therefore
    we expect no qualitative difference using the single data point versus using a full
    likelihood with proper treatment of uncertainties.

The MCMC results are summarized  in Table \ref{table:mcmc}.
Again, we find no detection of the dm-drf interaction
using the joint dataset {\tt PlanckTP+Lensing+DES},
though it is more constraining than another joint dataset {\tt PlanckTP+PlanckSZ}
with the CCCP prior.

\section{Lyman-$\alpha$ forest data}
\label{sec:lyman}

Ly$\alpha$ forest observations have been used as a cosmological probe
for the past two decades \cite[e.g.][]{Croft1998, McDonald2000, McDonald2006}.
Ly$\alpha$ absorption is sensitive to the density of neutral gas
in a relatively low-density, smooth environment, which
is tightly correlated with the underlying dark matter density on large scales.
Many of these observational results are based on a direct measurement of
the Ly$\alpha$ forest power spectrum $P_F(k)$, a statistical property
of the transmitted flux fluctuations
\be
\delta_F(\lambda) = e^{-\tau(\lambda)}/\langle e^{-\tau(\lambda)} \rangle - 1,
\ee
where $\lambda$ is the observed wavelength of Ly$\alpha$ emission,
and $\tau$ is the optical depth to Ly$\alpha$ absorption.
The tight correlation between the neutral gas density
and the underlying dark matter density
allows a determination of the matter power spectrum from the
Ly$\alpha$ forest power spectrum $P_F(k)$.
For this purpose, hydrodynamical simulations are required
to compute $P_F(k)$ for a given
initial linear matter power spectrum $P_L(k, z_i)$ at
some high redshift $z_i$,
due to the complexities in the non-linear evolution of dark matter
and hydrodynamical processes.

\begin{figure}
\includegraphics[scale=0.45]{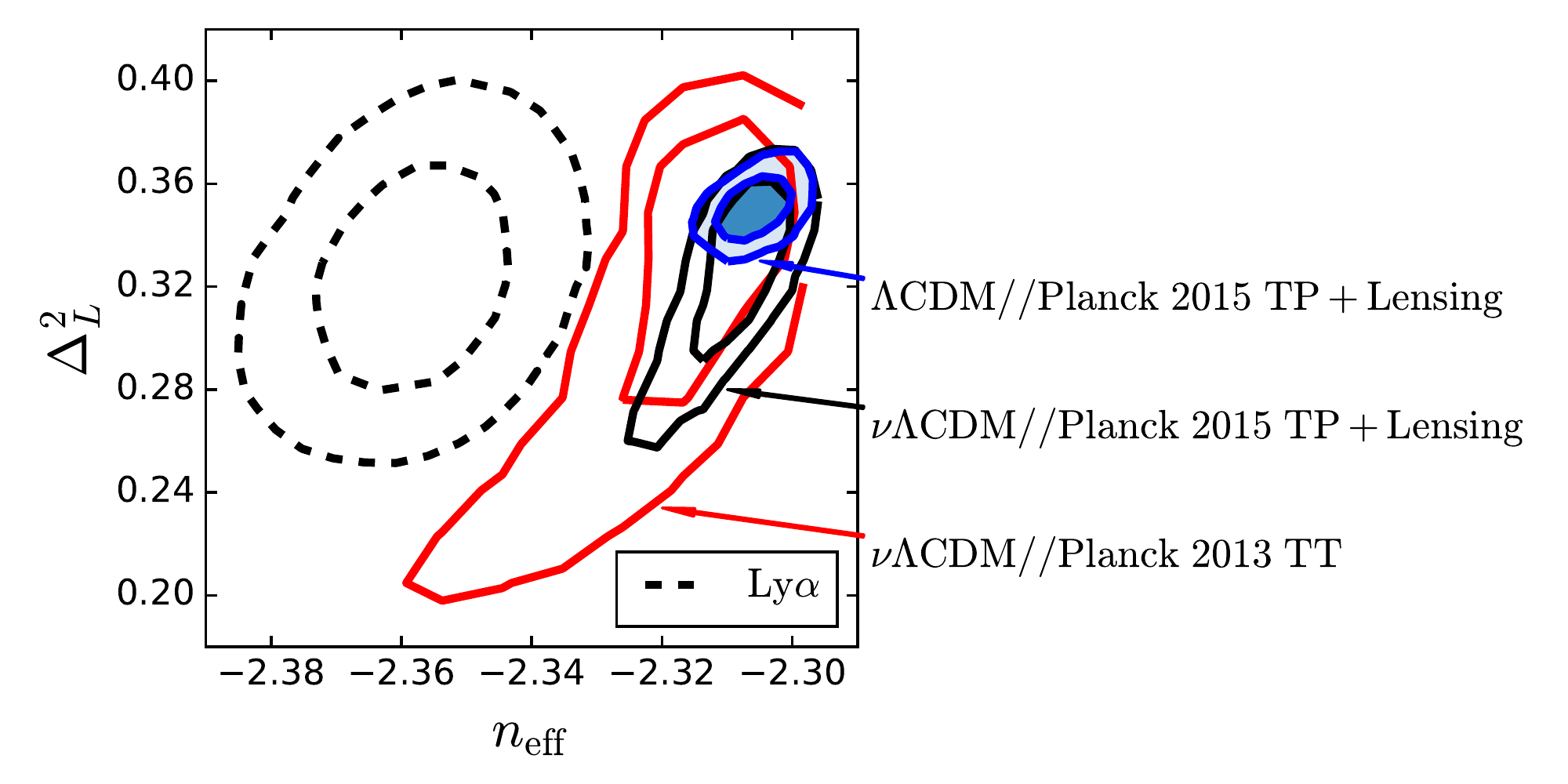}
\caption{\label{fig:lyalpha_tension}
The tension between Ly$\alpha$ and CMB in $\Lambda$CDM cosmology with
neutrino mass fixed as $0.06$ eV ($\Lambda$CDM) or varying ($\nu\Lambda$CDM)
where $\Delta_L^2$ and $n_{\rm eff}$ are the amplitude and the slope of the
linear matter power spectra at $k\simeq h \ {\rm Mpc}^{-1}$ and at $z=3$.}
\end{figure}

Compared with CMB data and LSS measurements including DES and {\sl Planck} SZ,
the latest measurements \cite{Palanque-Delabrouille2013,Palanque-Delabrouille2015c,Palanque-Delabrouille2015b}
of the Ly$\alpha$ forest flux power spectrum from the Baryon Oscillation Spectroscopic Survey
extends sensitivity to the matter power spectrum to smaller scales.
The constraints on the amplitude  $\Delta_L^2 = k^3 P_L(k,z)/2\pi^2$
and the slope $n_{\rm eff} = d\ln P_L(k,z)/d\ln k$
at $k = 0.009 ({\rm s/km})\times H(z)/(1+z)$ and  $z=3$
are explicitly given in \cite{Palanque-Delabrouille2015c},
where $H(z)$ is the Hubble expansion rate.
As pointed out in \cite{Minor2015}, the matter power spectrum
derived from the Ly$\alpha$ forest data yields a
comparable amplitude but a much steeper slope at scale $k\sim {\rm Mpc}^{-1}$,
compared with those derived from \textsl{Planck} CMB data, assuming $\Lambda$CDM.
We plot these constraints in Figure \ref{fig:lyalpha_tension},
which clearly shows that inferences of the matter power spectrum
from the Ly$\alpha$ forest data are highly inconsistent with the \textsl{Planck} CMB data,
assuming $\Lambda$CDM. The discrepancy has increased from the first release of
\textsl{Planck} data to the second, since the second yields
a flatter slope $n_{\rm eff}$ with a reduced uncertainty
(likely due to a larger $n_{\rm s}$ and a tighter constraint
on $\omega_{\rm m}$ \cite{PlanckCollaborationXIII2015})
Allowing neutrino mass to vary does not do much
to reconcile the discrepancy in the matter power slope $n_{\rm eff}$.

\begin{figure}
\includegraphics[scale=0.6]{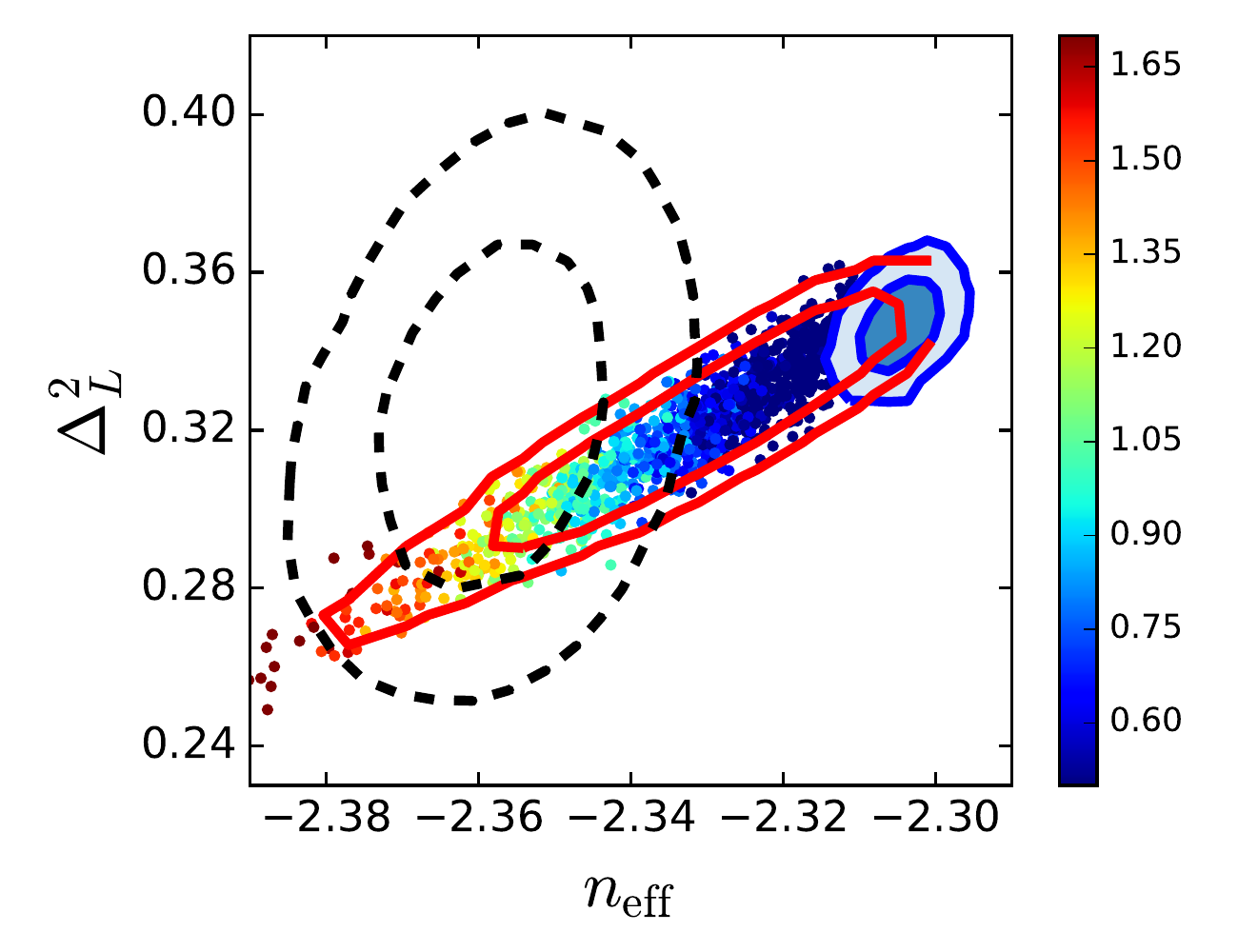}
\caption{\label{fig:lyalpha_dmdrf}
Comparison of the Ly$\alpha$ constraints
on the amplitude $\Delta_L^2$
and the slope $n_{\rm eff}$ of the matter power spectrum with those derived from
$\Lambda$CDM (blue filled contours) and
the dm-drf interaction model (red unfilled contours) using {\tt PlanckTP+Lensing+DES},
where the color points denote different $\Gamma_0$ values in unit of $10^{-7} {\rm Mpc}^{-1}$.}
\end{figure}

The steeper slope $n_{\rm eff}$ derived from Ly$\alpha$ data at scale $k\sim {\rm Mpc}^{-1}$
implies a scale-dependent matter power suppression which aligns well with
the dm-drf interaction picture.
To examine whether the Ly$\alpha$ data is in agreement with other datasets
in the dm-drf interaction model, we plot the $\Delta_L^2 - n_{\rm eff}$ contours
derived from {\tt PlanckTP+Lensing+DES} in Figure \ref{fig:lyalpha_dmdrf}.
We see that the joint dataset favors the dm-drf interaction model
(with interaction rate $10^7\Gamma_0/{\rm Mpc}^{-1}$ in the range of $[0.9, 1.6]$).

Our results serve to highlight the potential importance of these inferences of the
matter power spectrum from the Ly$\alpha$ data. If they are substantially free from bias
and have adequately captured all significant sources of uncertainty, then the discrepancy
with the Planck-conditioned predictions of $\Lambda$CDM are extremely interesting. Possible solutions
to this discrepancy include the dark matter model we are studying here, as well as a negative running
$d n_{\rm s}/d\ln k$ \cite{Minor2015,Palanque-Delabrouille2015b}
or possibly a different dm-drf interaction.

\section{Generalized dm-drf models}
\label{sec:general}

\begin{figure*}
\includegraphics[scale=0.6]{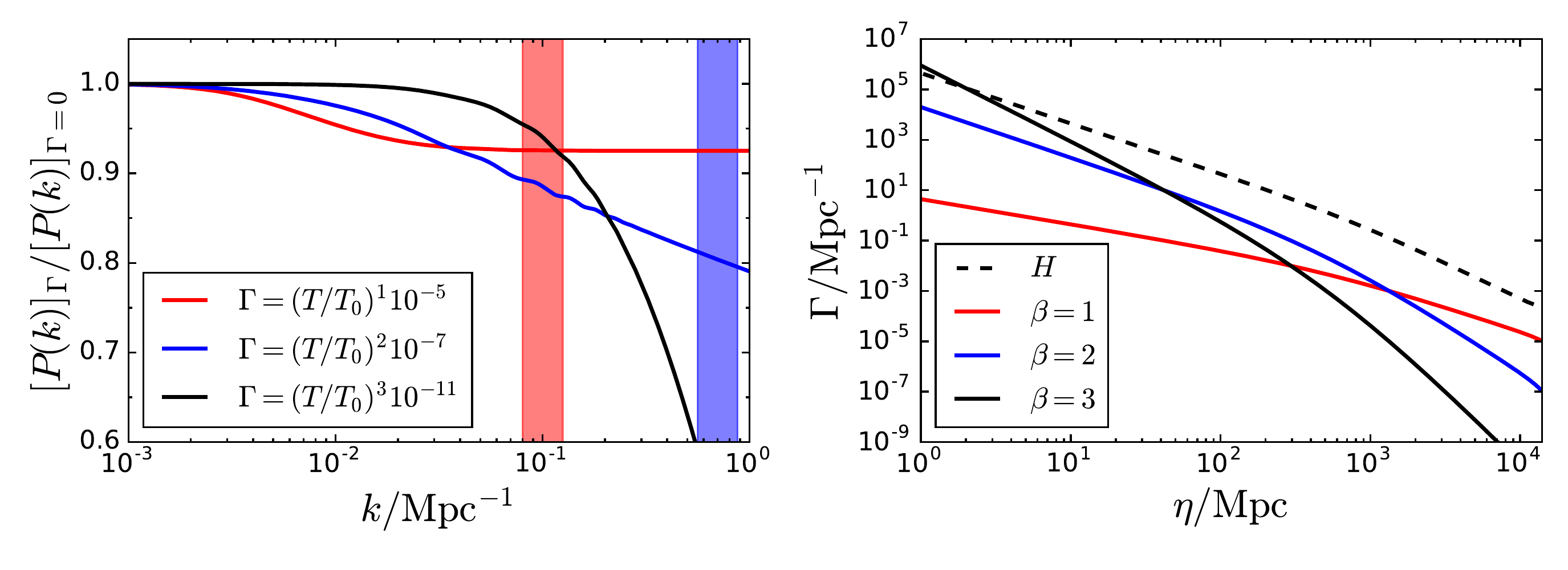}
\caption{\label{fig:beta123}
Upper Panel: the matter power spectrum suppression from
different dm-drf interaction rates $\Gamma\propto T^\beta$ $(\beta=1,2,3)$,
where the red band denotes the modes $\sigma_8$ is sensitive to
and the blue band denotes the modes Ly$\alpha$ measurement is sensitive to.
Lower Panel: the comparison between the Hubble expansion rate
and the dm-drf interaction rates.}
\end{figure*}

\begin{figure*}
\includegraphics[scale=0.43]{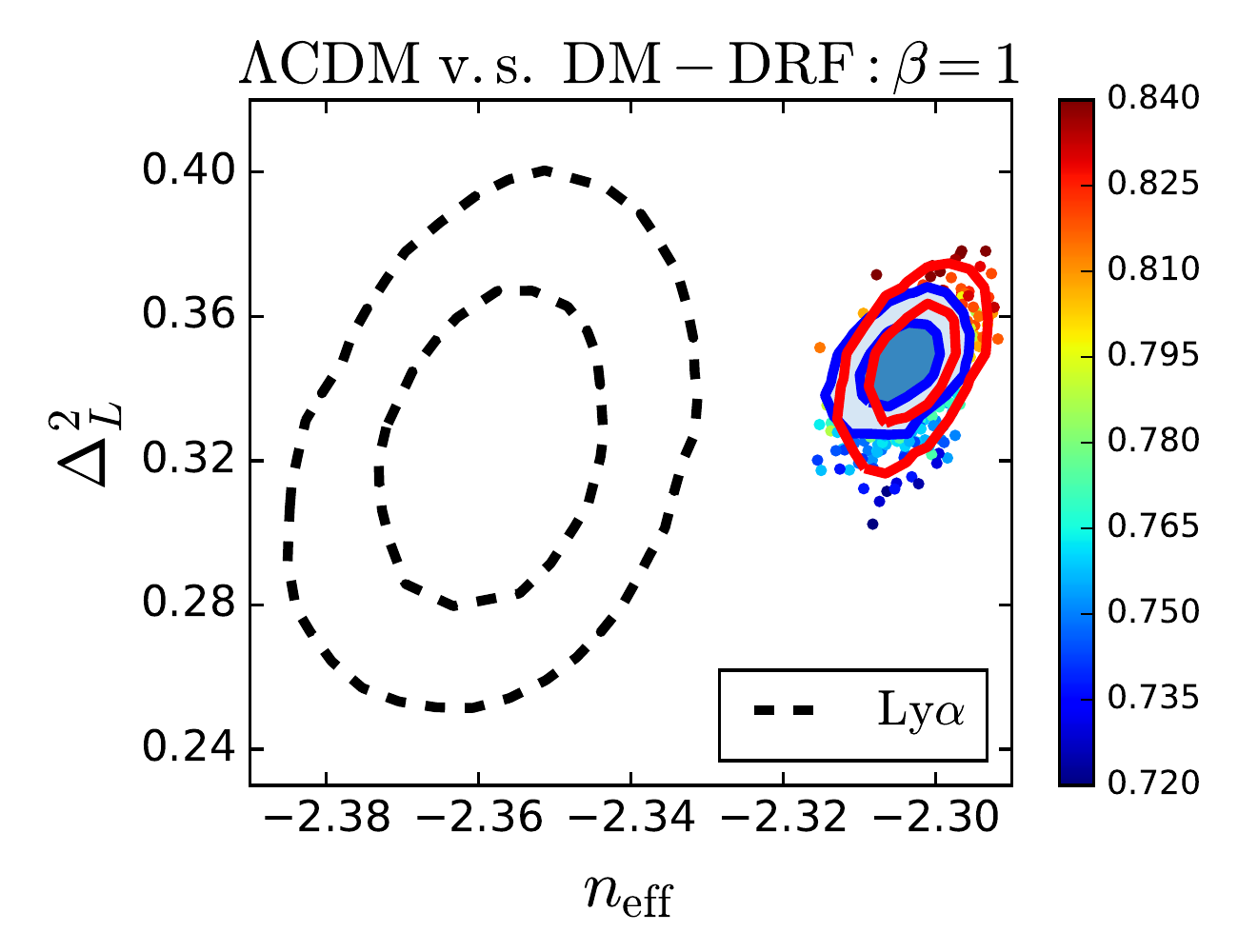}
\includegraphics[scale=0.43]{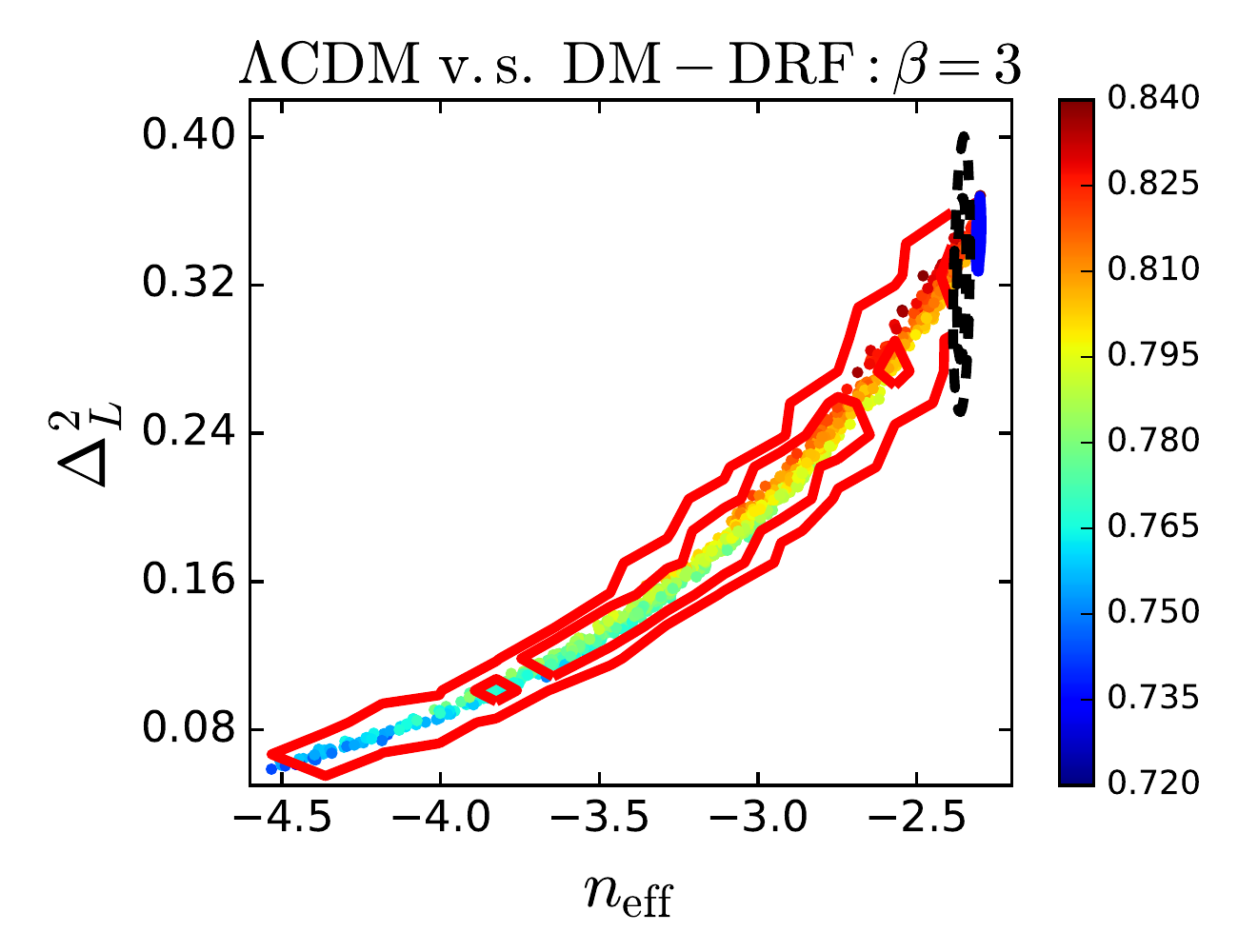}
\includegraphics[scale=0.43]{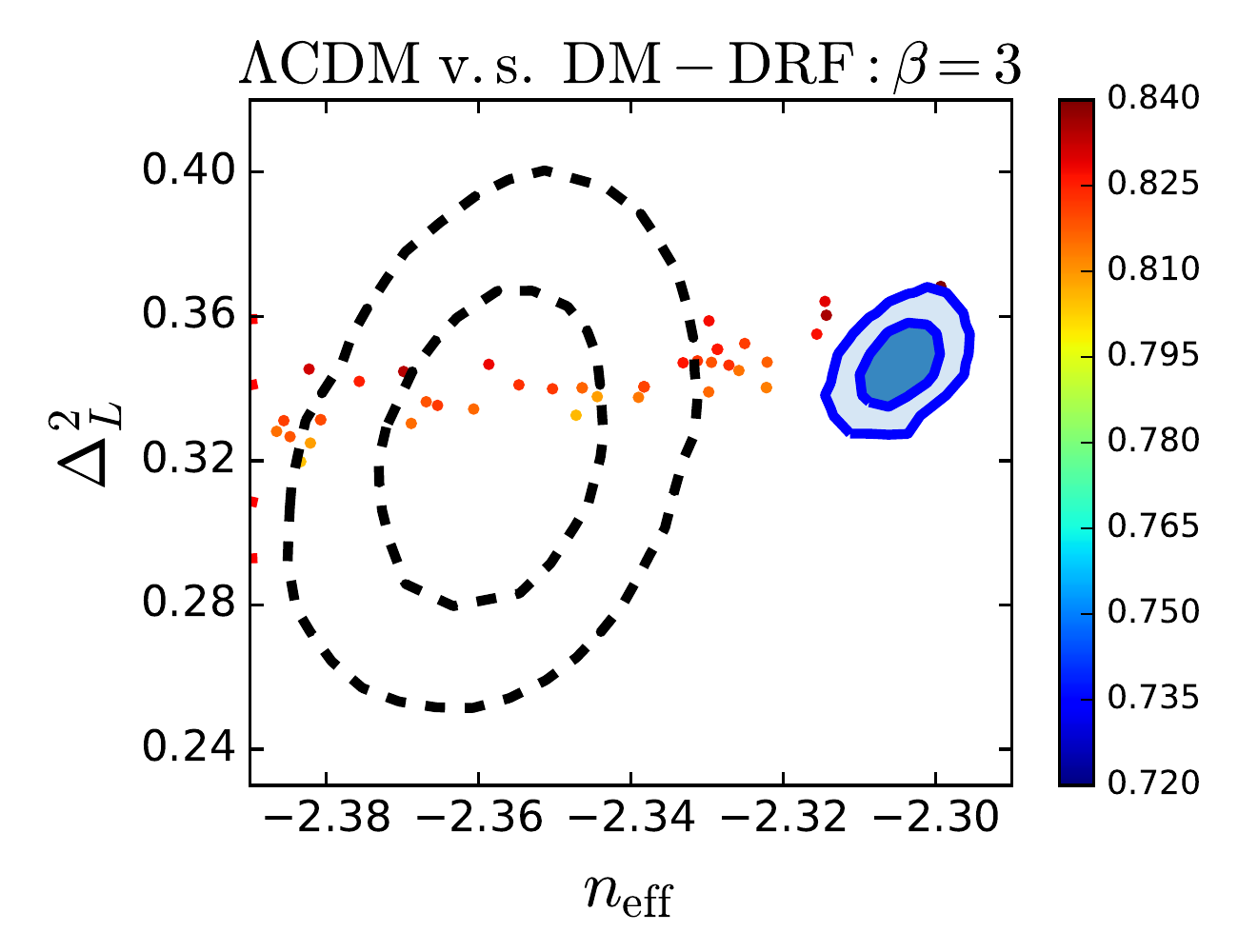}
\caption{\label{fig:lyalpha}Comparison of the Ly$\alpha$ constraint
on the amplitude $\Delta_L^2$
and the slope $n_{\rm eff}$ of the matter power spectrum with those derived from
$\Lambda$CDM and dm-drf interaction models using {\tt PlanckTP+Lensing+DES},
where the blue filled contours are the results of the $\Lambda$CDM cosmology,
the red unfilled contours are the results of the dm-drf models,
and the color bars denote $\sigma_8$ for dm-drf with $\beta=1,3$, respectively.}
\end{figure*}

In this section, we extend our exploration from the canonical dm-drf model
with interaction rate $\Gamma=\Gamma_0(T/T_0)^2$ to generalized models
with interaction rate $\Gamma=\Gamma_0(T/T_0)^\beta$ ($\beta = 1,2,3$).
We first briefly discuss the imprints  of the generalized dm-drf interaction rates
on the CMB power spectra and the matter power spectrum,
then constrain these models using CMB data and LSS measurements.

The imprints of the three different models on the matter
power spectrum suppression are pretty distinct at small scales,
as shown in Figure \ref{fig:beta123}.
For $\beta=3$, dark matter and dark radiation are tightly coupled deep in the
radiation-dominated era. Dark matter perturbations oscillate instead of
growing, therefore all modes entering the horizon when $\Gamma\gtrsim H$
are strongly suppressed.
In the $\beta=1$ case, the interaction is negligible at early time.
For sufficiently small modes entering horizon early,
the power suppression is dominated by late time when the dm-drf interaction becomes important.
For large modes entering horizon when $\rho_{\rm drf}/\rho_{\rm dm}$ becomes vanishingly small,
the dark matter overdensity growth is unaffected by the interaction.
For the intermittent modes, the power suppression
is determined by several factors, including the coupling strength today $\Gamma_0$,
the matter-radiation equality where the $\Gamma/H$ dependence on the scale factor changes,
and the dark matter and dark radiation energy density ratio $\rho_{\rm drf}/\rho_{\rm dm}$.
Therefore we see a power suppression plateau on the small-scale end, no suppression on
the large-scale end, and a smooth transition in between.
Different from the $\beta=2$ case, both $\beta=1$ and $\beta=3$ interactions
introduces new length scales to the matter power suppression.

 Following the argument given in Section \ref{subsec:impacts_cmb}, it
 is not hard to figure out the impacts of the general interaction rates on
 the temperature and polarization power spectra.
 For example, we expect that the $\beta=3$ interaction tends to suppress the amplitudes
 of large $k$ modes entering the horizon at radiation domination
 and when $\Gamma/H$ is noticeable, while
 leaving no imprint on the amplitudes of small $k$ modes entering the horizon when $\Gamma/H$
 is negligible;  on the contrary, the $\beta=1$ interaction should only affect
 large $k$ modes. We have modified  {\tt CAMB} to allow for all the three interaction models,
 and numerical results confirm our qualitative expectations above.
 We find that the imprints of the three different models are too subtle to
 be distinguished via \textsl{Planck} CMB data, so we do not plot them here.

We constrain the two models using the joint dataset {\tt PlanckTP+Lensing+DES}, finding no detection of interaction for either of the two new cases.
Similar to previous section, we also examine whether these two models reconcile the
Ly$\alpha$-CMB tension. As shown in Figure \ref{fig:lyalpha},
the $\beta=1$ interaction does not change the amplitude and the slope much,
and the $\beta=3$ interaction leads to an overwhelming suppression.
We see that neither of the two help to reconcile the Ly$\alpha$-CMB tension.

\section{Summary}
\label{sec:summary}
In this paper, we reinvestigated  the non-Abelian dark sector model proposed
by \cite{Buen-Abad2015}.
We examined the impact of the dm-drf interaction on the CMB power spectra and the
matter power spectrum in detail. We found that the dm-drf interaction affects the amplitudes of CMB power spectra by modifying the gravitational potential decay, but only slightly.
We verified the presence of a logarithmic suppression in the matter power spectrum that
originates from the self-similar suppression of the matter overdensity.
We also constrained the dm-drf model using CMB and LSS measurements in a more systematic way.

We found that \textsl{Planck} SZ plays the key role in the previously claimed
detection of dm-drf interaction. However the SZ cluster counts constraint
is limited by uncertainty in the cluster mass scale determination, which is usually parametrized as the mass bias parameter $b$.
We confirmed the $3 \sigma$ detection using the \textsl{Planck} CMB data
and the SZ data fixing the bias parameter to be constant, $1-b=0.8$, as
done in previous analyses. But, when we included uncertainties in $1-b$, the detection of dm-drf interaction essentially disappeared.

We also show that the latest inferences of the matter power spectrum
from Ly$\alpha$ forest data are highly inconsistent with the \textsl{Planck} CMB data,
assuming $\Lambda$CDM, and that the joint data sets favor a non-zero dark sector interaction.
Thus if these matter power spectrum inferences are free from significant systematic error, and if the reported uncertainties accurately include all sources of uncertainty, these data are more sensitive to the impact of dm-drf interactions and provide us with a significant detection. Even so, there are other possible ways to reconcile the Planck and Ly$\alpha$ forest data  such as a non-zero running of the scalar spectral index
$d n_{\rm s}/d \ln k$ \cite{Minor2015,Palanque-Delabrouille2015b}.

We also explored two different phenomenological dm-drf interaction models characterized by interaction rates scaling with temperature in different power laws, and found neither of these interactions is favored by current CMB and LSS data.

We are unsure what to make of these inferences of the matter power spectrum from
the Ly$\alpha$ forest data. We hope our work serves to motivate
further study of these data. Were a different group to reach similar
conclusions independently, even if from the same data, that would bolster our
confidence. Another avenue for progress is measurements that can decrease uncertainty
in $1-b$ as the cluster mass function has the statistical power to make a
detection absent that uncertainty, if the interaction strength is at the higher end of the range consistent with Ly$\alpha$ data.

\begin{acknowledgements}
    We thank Julien Lesgourgues for helpful comments on an earlier version of the manuscript.
    ZP thanks Lachlan Lancaster for his valuable help in modifying {\tt CAMB}.
    ZP is supported by the UC Davis Dissertation Year Fellowship.
    MK is supported by the National Science Foundation Grant PHY-1620638.
    This work made extensive use of the NASA Astrophysics Data
    System and of the astro-ph preprint archive at arXiv.org.
    All the computation and plots are done with the Boltzmann codes
    {\tt CAMB} and {\tt CLASS}, and MCMC codes {\tt Cosmomc}, {\tt MontePython}
    and  {\tt Cosmoslik}.
\end{acknowledgements}

\bibliography{ms,mk}

\end{document}